\documentclass[conference]{IEEEtran}
\IEEEoverridecommandlockouts
\usepackage{cite}
\usepackage{amsmath,amssymb,amsfonts}
\usepackage{algorithmic}
\usepackage{graphicx}
\usepackage{textcomp}
\usepackage{xcolor}
\usepackage{soul}  
\usepackage{booktabs}		

\usepackage{comment,svg,subfigure}
\def\BibTeX{{\rm B\kern-.05em{\sc i\kern-.025em b}\kern-.08em
    T\kern-.1667em\lower.7ex\hbox{E}\kern-.125emX}}
\begin{document}

\title{Optimal Coordination of Local Flexibility from Electric Vehicles  with Social Impact Consideration\\

}
\author{\IEEEauthorblockN{Si Chen$^{1}$, Benoit Couraud$^{1}$, Sonam Norbu$^{1}$, Merlinda Andoni$^{1}$, Zafar Iqbal$^{2}$, Sasa Djokic$^{2}$, Desen Kirli$^{2}$, \\ Satria Putra Kanugrahan$^{1}$, Paolo Cherubini$^{3}$, Susan Krumdieck$^{3}$,  Valentin Robu$^{4}$, David Flynn$^{1}$}
	
	\IEEEauthorblockA{$^{1}$James Watt School of Engineering, University of Glasgow, Glasgow (UK), \\
		$^2$ University of Edinburgh, Edinburgh (UK), 
		$^3$ Heriot-Watt University, Stromness (UK), \\
        $^4$ Intelligent and Autonomous Systems Group, CWI, Amsterdam (The Netherlands)\\
        }
}
\maketitle

\begin{abstract}
The integration of renewable energy sources (RES) and the convergence of transport electrification, creates a significant challenge for distribution network management e.g. voltage and frequency violations, particularly in rural and remote areas. This paper  investigates how smart charging of electric vehicles (EVs) can help reduce renewable energy curtailment and alleviate stress on local distribution networks. We implement a customised AC Optimal Power Flow (AC OPF) formulation which integrates into the optimisation an indicator reflecting the social impact of flexibility from EV users, based on the analysis of historical EV charging behaviours. The contribution of EV owners to reducing wind curtailment is optimised to enhance the acceptability of flexibility procurement, as the method targets EV users whose charging habits are most likely to align with flexibility requirements.
Our method integrates social, technological, and economic perspectives with optimal flexibility coordination, and utilises clustering of EVs through a k-means algorithm. To ensure scalability, we introduce a polar coordinate-based dimension reduction technique. The flexibility optimisation approach is demonstrated on the Orkney grid model, incorporating demand and wind farm generation data, as well as multi-year charging data from 106 EVs. Results indicate that, by building upon the existing habits of EV users, curtailment can be reduced by 99.5\% during a typical summer week—the period when curtailment is most prevalent. This research demonstrates a foundational and transferable approach which is cognisant of social-techno-economic factors towards accelerating decarbonisation and tackling the stochastic challenges of new demand and generation patterns on local distribution networks.

\end{abstract}


\section{Introduction}
Decarbonising energy usage requires strong efforts to integrate renewable energy sources (RES) and electrify high-emission demands such as heating and, more importantly, transport, which is the main source of greenhouse gas emissions in the UK~\cite{DESNZ2024GHG}.
However, these transformations introduce new challenges for distribution networks, which must accommodate increasingly large and dynamic patterns of energy production and consumption~\cite{ray2023review}.
In response, smart grid solutions have been adopted~\cite{hasan2025framework,mokhtar2021prediction}, along with zonal segregation of distribution networks to mitigate constraints through Active Network Management (ANM) schemes~\cite{9281406}. In the Orkney Islands, for instance, substantial investments in wind energy have resulted in RES installation over the available grid capacity. As a result, the ANM system operated by the distribution system operator (DSO) enforces curtailment of the most recently connected wind turbines~\cite{almoghayer2022integration}. Notably, many of these newer wind turbines are community-owned, amplifying the social impact of curtailment.
Simultaneously, the electrification of transport is driving an increase in the number of electric vehicles (EVs), which significantly raises electricity demand and introduces further technical challenges, including voltage imbalances, harmonic distortion, transformer and line overloading, and other operational and power quality issues~\cite{srivastava2023electric}. However, EVs also offer potential as residential flexible loads that could provide demand response and mitigate renewable generation curtailment~\cite{couraud2023responsive}. Yet, this potential introduces concerns over the technical feasibility and social acceptability of residential flexibility. It is therefore essential to develop technical solutions that enable flexibility from EV charging while minimising disruption for end users.

To address this issue, the present study proposes a novel AC Optimal Power Flow (AC OPF) approach that incorporates empirical insights from historical EV charging data. EV users charging habits are  integrated into the power flow optimization as a social impact factor—reflecting the extent to which users’ charging preferences align with the flexibility profile needed to reduce renewable energy curtailment. 
As analysing individual profiles at scale is impractical, we adopt a clustering approach to represent charging behaviour patterns. While techniques such as k-means, decision tree-based clustering, self-organising maps, and hierarchical clustering have been used in previous studies~\cite{ray2023review, hasan2025framework}, most do not explore how such clusters can be directly used by grid operators, nor address the challenge of scalability for future large-scale EV deployment. To address these issues, this paper introduces a novel EV charging profile classification method using a dimension reduction technique that improves scalability. The proposed classification and novel flexibility optimisation approach have been validated in simulations based on the actual Orkney grid and demonstrated a significant curtailment reduction. 

The remainder of this paper is structured as follows: Section~\ref{sec:clustering} details the methodology used to classify EV charging patterns for standard and proposed k-means clustering approaches. Section~\ref{sec:acopf} presents the novel AC OPF-based coordination framework, which leverages EV user clusters to optimise grid services and reduce curtailment, while Section~\ref{sec:conclusion} concludes this work.

\section{EV Charging Behaviour Clustering}
\label{sec:clustering}
In this paper, we develop a method that utilises flexibility from EV users in order to reduce wind curtailment, ultimately benefiting the local community. This flexibility involves increasing the total charging demand of EVs during periods
when curtailment is anticipated, which is realised by connecting/shifting more EVs to charge at these times.
However, studies have shown that residential users are less inclined to provide flexibility through their EVs compared to other appliances, due to concerns that their vehicles may not be sufficiently charged when needed~\cite{SRIDHAR2023121204}. To address this issue, we propose analysing EV users’ charging behaviours to selectively target those whose typical charging times align with periods when flexibility is required. For example, EVs that typically charge at night would not be asked to increase their charging during the day.
Analysing each EV individually, however, is neither efficient nor scalable. Therefore, we group EVs into clusters based on similar charging behaviours. To achieve this, we adopt an unsupervised learning approach using the k-means clustering algorithm to identify common usage patterns among users. We initially outline a standard k-means clustering implementation, and then propose an approach that ensures greater scalability for large-scale EV data.

\subsection{Standard k-means approach}
The k-means clustering algorithm is a centroid-based technique that uses an iterative process to determine $k$ centroids, each representing a cluster of EV charging profiles where members exhibit similar behaviour patterns~\cite{hasan2025framework}. Given a dataset of $n$ EV users, we define $n$ average EV charging profiles, denoted as $x_j \ (1 \leq j \leq n)$, where each profile captures the user's average charging power over the observed period.
For example, in a dataset with minute-level data from January 2020 to January 2021, each user has 365 daily load profiles, each consisting of 1,440 data points ($60 \cdot 24$). At each time step, the value is either zero (if the EV is not charging) or the corresponding charging power. By averaging these 365 daily curves, we obtain a single representative profile for each user—referred to as their average EV profile.
These $n$ average profiles are then assigned to the nearest centroid, where each centroid is the mean profile of all members in its respective cluster. The assignment and centroid update process is repeated iteratively to minimise the total sum of Euclidean distances between each profile and its cluster centroid. This objective is formally expressed in Eq.~\ref{eq:kmeans}:

\begin{equation}
J = \sum_{i=1}^{k} \sum_{x_j \in C_i} \left\| x_j - \mu_i \right\|^2
\label{eq:kmeans}
\end{equation}
where $k$ is the number of clusters considered, $C_i$ is the set of EV users in the $i$-th cluster, $x_j$ is a data point in cluster $C_i$, and $\mu_i$ is the centroid (mean) of cluster $C_i$. The optimal number of clusters can be determined using both the elbow and silhouette methods.\newline
\begin{figure}[h!]
\centering
\vspace{-5mm}
\includegraphics[width=1\columnwidth]{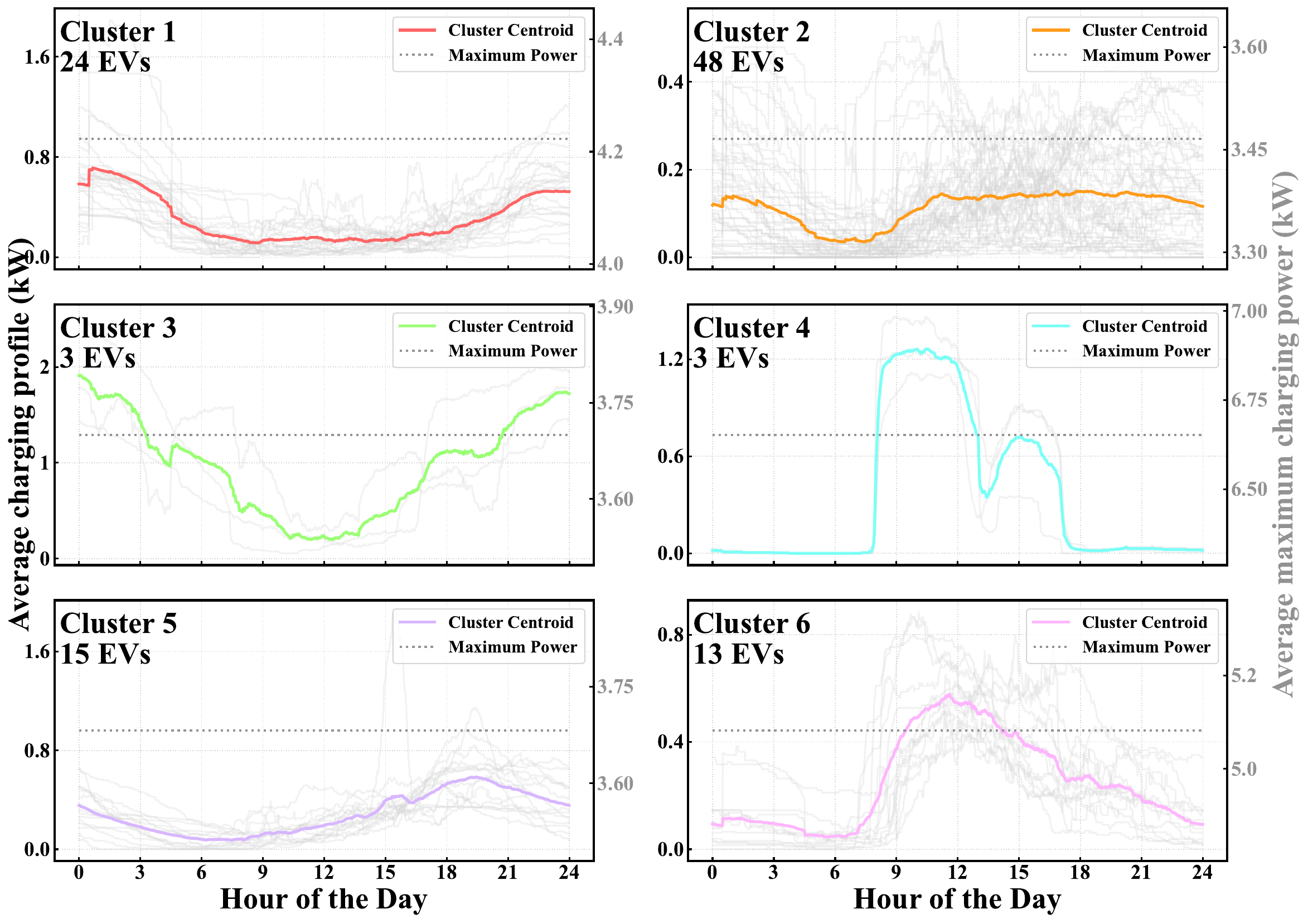}
\caption{Standard k-means clustering using EV profiles for week days.}
\vspace{-5mm}
\label{fig:KmeanClustering}
\end{figure}

In this subsection, we outline a standard k-means approach, which serves as the baseline, where each vector $x_j$ (array of 1,440 elements) corresponds to the average EV power consumption at a specific minute of the day, computed across the entire dataset period. We applied this method to a dataset of 106 EV users~\cite{couraud2023responsive}, whose charging events were recorded between 2020 and 2024, resulting in over 26,000 individual charging sessions.
Using both the elbow and silhouette methods, we identify $k = 6$ optimal number of clusters with a total intra-cluster Euclidean distance of 5.7 kW across all users. Fig.~\ref{fig:KmeanClustering} presents the centroid charging profiles for each cluster, alongside the average EV charging profiles of the cluster members (primary vertical axis). It also shows the average maximum charging power for each EV as a dotted line (secondary vertical axis), calculated over charging periods only, and excluding the top 10th percentile to reduce the influence of outliers.
The left-hand scale of Fig.~\ref{fig:KmeanClustering} corresponding to the coloured curves, represents the average EV charging profile. Lower values on this scale indicate users who charge their EVs less frequently.
In alternative implementations, the vector $x_j$ may include features beyond the average charging profile, such as the average charging power, the ratio of weekend to weekday charging, and the proportion of days with charging activity versus inactivity. Some studies attempt to reduce the dimensionality of $x_j$ further by considering only a few features—such as the typical start time of charging sessions and the total connection duration~\cite{singh2022smart}. However, such simplified representations often fail to capture users with multiple charging events in a day—a behaviour commonly observed in our dataset. Consequently, these reduced-dimension approaches were not adopted in this study.
Based on the clustering results, we identified several distinct charging patterns, summarised in Table~\ref{tab:clusterDescription}. A scalable k-means method is shown in the next subsection.

	\begin{table}[ht]
		\caption{Description of the charging patterns cluster.} 
		\label{tab:clusterDescription}
		\small 
		\centering 
		\begin{tabular}{lc} 
			\toprule[\heavyrulewidth]\toprule[\heavyrulewidth]
			\textbf{Cluster}  & \textbf{Pattern description} \\ 
			\midrule
			Cluster 1  &  Charging at home, at night \\
			Cluster 2  &  Rarely charged, mostly at night and afternoon\\
			Cluster 3  &  Often charged at home in the night\\
			Cluster 4  &  Often charged in the day, at higher power\\
			Cluster 5  &  Rarely charged, mostly in the evening\\
			Cluster 6  &  Charging during the day\\
			\bottomrule[\heavyrulewidth] 
		\end{tabular}
	\end{table}

\subsection{Polar coordinates based approach}
\begin{figure}[tbp]
\centering
\includegraphics[width=1\columnwidth]{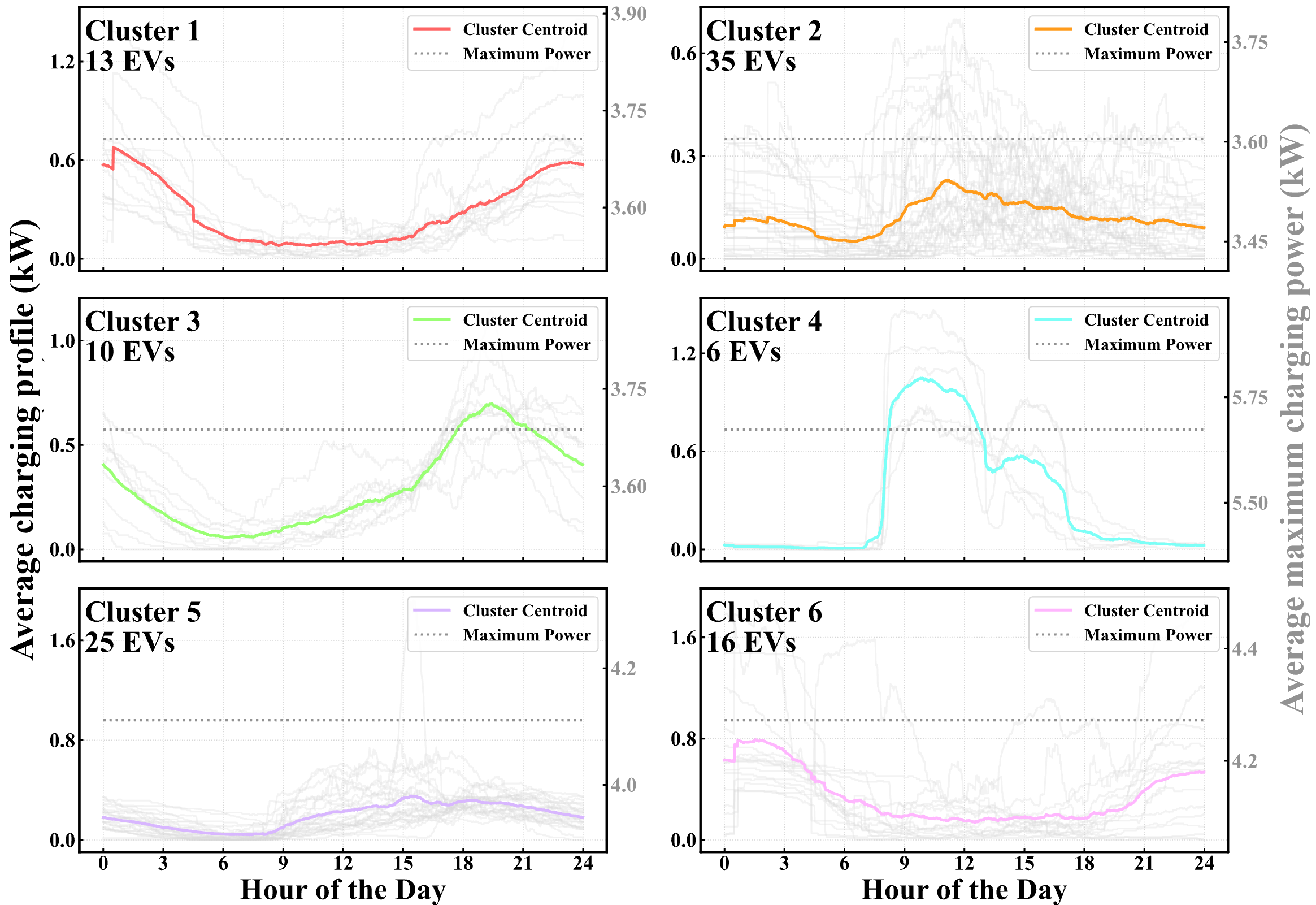}
\vspace{-4mm}
\caption{Proposed k-means clustering using polar coordinates.}
\vspace{-3mm}
\label{fig:KmeanNEW}
\end{figure}

While running the k-means algorithm is computationally efficient for a dataset of 106 EV users, each with 1,440 dimensions, the process could become significantly more resource-intensive in the future when scaling to the whole fleet of EVs in Orkney after electrification of transport. 
To address this scalability challenge, we propose a method that transforms charging information into polar coordinates. In this approach, the original high-dimensional vector $x_j$ is replaced with a simplified 2-dimensional vector $(x_{i_1}, x_{i_2})$, as defined in Eq.\ref{eq:kmeans_x} and Eq.\ref{eq:kmeans_y}.

\begin{equation}
x_{i_1} = \sum_{t=0:00}^{24:00} \left[ \sin \left(2\pi \cdot \dfrac{t}{24}\right)\cdot P_i(t) \right]
\label{eq:kmeans_x}
\end{equation}
\begin{equation}
x_{i_2} = \sum_{t=0:00}^{24:00} \left[ \cos \left(2\pi \cdot \dfrac{t}{24}\right)\cdot P_i(t) \right]
\label{eq:kmeans_y}
\end{equation}
where $t$ is the time of the day, $P_i(t)$ is the average charging power of the $i$-th EV at time $t$, as computed in the previous section for the average EV load profiles.
This vector captures the EV charging power, and using polar coordinates allows for representing temporal variations in the profiles. The results of applying the proposed k-means clustering method with these polar coordinates are shown in Fig.~\ref{fig:KmeanNEW}. The sum of Euclidean distances is $J = 6.5$ kW, which, although not as optimal as the standard method, is achieved with a significant reduction in dimensionality, thereby improving scalability. The clusters obtained using this approach are comparable to those from the traditional method, except for cluster 6, with the added benefit of a more uniform distribution of EVs across the clusters than what was obtained with the first method—which led to 2 clusters with 3 EVs only.

\section{Optimal Coordination of EV Flexibility for Grid Management}
\label{sec:acopf}
Once EV users have been categorised, this information can be used by an aggregator or the Distribution System Operator (DSO) to target individual EV users whose charging patterns align with the flexibility required to mitigate anticipated grid constraints. For instance, if a network constraint is anticipated in the afternoon of a particular day, the goal is to request flexibility from EV users who will not be charging at the time of curtailment, but who are typically categorised as more likely to charge their vehicles during afternoons. In addition to this time-based correlation, it is also essential to target those users whose flexibility contributions will have a tangible impact on the specific grid issue—whether that be curtailment, line overloading, or voltage excursions.
To achieve this, we have developed an AC Optimal Power Flow (AC OPF) formulation that integrates the likelihood of EV users ability to provide flexibility. This formulation builds on the AC OPF model from~\cite{10863672}, incorporating a semidefinite programming (SDP) convex relaxation. The constraints are structured using a branch flow model, and the optimisation includes a multi-objective function that determines the optimal set of flexible assets to be contracted in order to resolve local grid issues such as voltage violations or network congestion. 

In summary, the objective of the optimisation function is to find the most cost-efficient and socially acceptable mix of flexibility from EVs ($\Delta P_{EV_{i,t}}^{+/-} \thickspace \forall$ EV users $i$) that can solve a network constraint. In this study, we focus on the increase of EV charging demand $\Delta P_{EV_{i,t}}^{+}$ to match oversupply from wind producers and reduce their risk of curtailment.  Furthermore, as individual EV user consideration is not practical, we aggregate them by the cluster to which they belong. All EV users that belong in cluster $C_i$ are considered for provision of flexibility $\Delta P_{EV_{C_i,t}}^{+}$, and the optimal quantity for every time step is derived by the minimisation of the objective function $f$:
\begin{equation} 
\begin{aligned}
	f = &  \sum_{i \in EV users} \left( \pi_{C_i,t} \Delta P_{EV_{C_i,t}}^{+} \right)  + \\
	& \sum_{g \in \mathcal(G) } \left( \pi_{g} \Delta P_{g,t}^{-} \right) + \dfrac{1}{M_t}\sum_{k, j \in \mathcal{N}}  \text{tr} \left[ \left| z_{kj}\ell_{kj,t}\right|  \right]    
\end{aligned}
\label{EqOptim}	
\end{equation}
\noindent where $z_{kj}$ is the $N_{kj} \times N_{kj}$ impedance matrix between buses $k$ and $j$; $\mathcal{N}$ is the set of buses; $M_t$ is a tunable coefficient for each time step that balances the trade-off between realistic power flows (through the minimisation of losses) and the minimisation of the social impact of flexibility~\cite{10863672}, defined  as a quantification of the effort required by the user to provide the required flexibility, as detailed below; $\mathcal{G}$ is the set of generators that can be curtailed; $\Delta P_{g,t}^{-}$ is the amount of curtailment required from generator $g$, as determined by the optimisation; and $\pi_{g}$ is the associated curtailment cost. Although wind turbines in Orkney are curtailed without compensation under their current non-firm connection agreement, their curtailment cost $\pi_{g}$ is modelled as virtually higher than social impacts to prioritise EV flexibility—reflecting the objective of reducing curtailment. Finally, $\ell_{kj,t} \in \mathbb{C}^{N \times N}$ is an $N \times N$ Hermitian matrix representing the current flowing from bus $k$ to bus $j$, defined as follows:
\begin{equation} 
	\ell_{kj,t} = I_{kj,t}I_{kj,t}^H.
\end{equation}	
\noindent where $I_{kj,t} \in  \mathbb{C}^{N}$ is the current flowing from bus $k$ to $j$, and  superscript $H$ represents the Hermitian transpose.
Finally, the specificity of our approach lies in the integration of EV users' charging habits, which is captured through the social impact coefficient $\pi_{C_i,t}$ which is defined as $\pi_{C_i,t}=\frac{1}{P_{avg_{C_i}}(t)+\epsilon}$, where $0<\epsilon <<1$ and $P_{avg_{C_i}}(t) = \mu_i(t)$ is the mean charging pattern of cluster $C_i$, computed in Section~\ref{sec:clustering}, and shown in coloured curves in Fig. \ref{fig:KmeanClustering}. We consider that the social impact for requesting an EV user in cluster $C_i$ to increase its consumption at time $t$, is inversely proportional to its tendency to charge at time $t$, as expressed by its cluster's centroid. For example, an EV user belonging in Cluster 1, where EVs are mostly charged during the night, will have a lower demand turn-up flexibility cost during night times than during the day.

The optimisation needs to adhere to the physical constraints of the distribution grid, namely the Ohm's law at every bus $j$:

\begin{equation} 
	\begin{aligned}
		V_{j,t} = V_{k,t} - \left( S_{kj,t}z_{kj}^H+z_{kj}S_{kj,t}^H\right)  + z_{kj}\ell_{kj,t}z_{kj}^H
		\label{Ohms_Law}
	\end{aligned}	
\end{equation}	
\noindent where $V_{j,t}$ is the hermitian voltage  matrix of bus $j$, and $S_{kj,t} \in \mathbb{C}^{N \times N}$ is a $N \times N$ complex matrix that represents the apparent power flowing from bus $i$ to bus $j$.

Similarly, the power balance constraints at every bus is given by: 
\begin{equation} 
	\begin{aligned}
		&\sum_{k:k\longrightarrow j} \text{diag}\left( S_{kj,t} - z_{kj}\ell_{kj,t} \right) + \sum_{g \in j } \left(  \Delta P_{g,t}^{-} \right) \\
		&- \sum_{\forall C_i} \Delta P_{EV_{C_i^j,t}}^{+}
		=  \sum_{h:j\longrightarrow h} \text{diag}\left( S_{jh,t} \right)
		\label{PowerBalance}
	\end{aligned}
\end{equation}
where $\Delta P_{EV_{C_i^j,t}}^{+}$ corresponds to the flexibility of EV users who belong in cluster $C_i$ and who are connected to bus $j$. 
Other constraints include the lines' maximal capacity  constraints, the voltages limits constraints, the positive semi-definite constraint, the convex relaxation, as explained in~\cite{10863672}, and finally a constraint on the maximum flexibility provided by EVs, defined for every cluster $C_i$ and bus $j$, as:
	\begin{equation} 
			 0\leq   \Delta P_{EV_{C_i^j,t}}^{+} \leq  n_i^{j, remain}(t)\cdot P_{max_{C_i}}
			\label{EqFlexConstraint}
	\end{equation} 
where $n_i^{j, remain}(t)=N_i^j - n_i^j(t)$ is the number of EVs of cluster $C_i$ in bus $j$ that are not yet charging, with $N_i^j$ the number of EV from $C_i$ allocated to bus $j$, $n_i^j(t)$ the number of EVs from $C_i$ that are charging at time $t$ in average, computed from the data of EVs that were clustered in $C_i$ , and $P_{max_{C_i}}$ is the average maximum charging power in Cluster $C_i$, as shown in Fig. \ref{fig:KmeanClustering} and Fig. \ref{fig:KmeanNEW} by the dotted horizontal line. 

\begin{figure}[tbp]
\centering
\includegraphics[width=0.5\columnwidth]{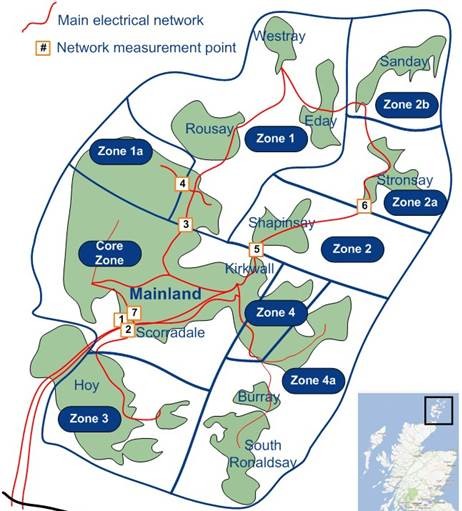}
\vspace{-2mm}
\caption{Orkney distribution system representation \cite{almoghayer2022integration}.}
\vspace{-3mm}
\label{fig:Orkney}
\end{figure}

\begin{figure}[bp]
\centering
\vspace{-5mm}
\includegraphics[width=1\columnwidth]{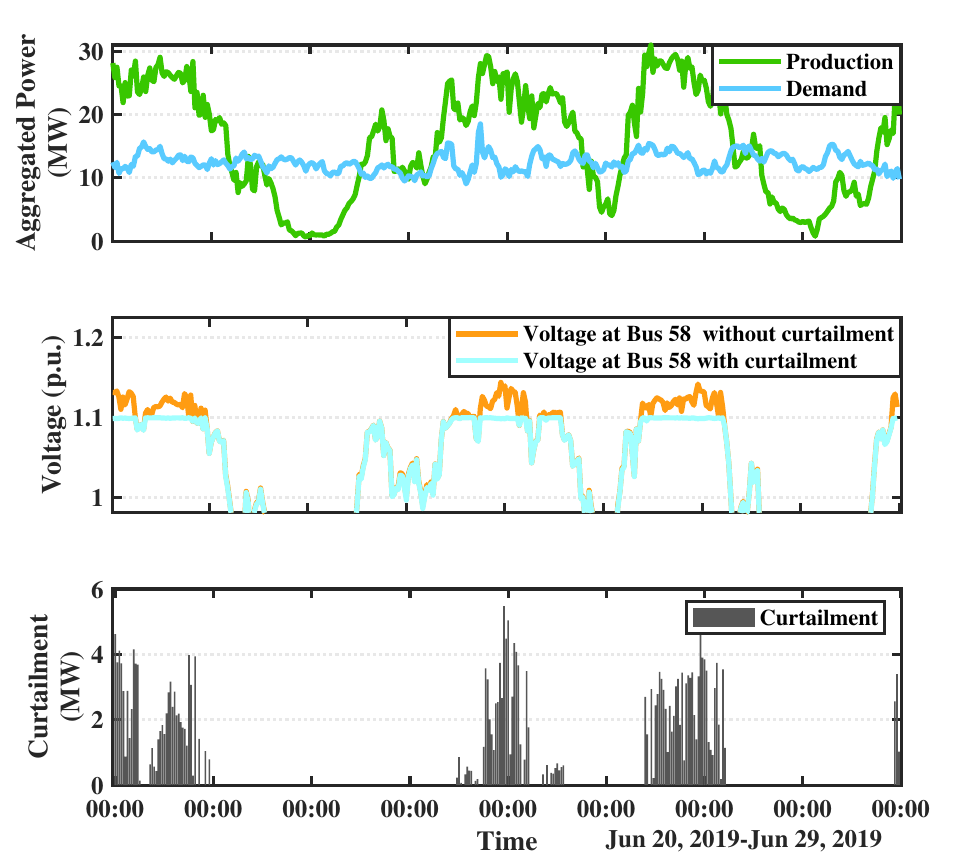}
\vspace{-5mm}
\caption{Curtailment events during the simulated period, with associated aggregated power production and consumption in the Orkney Islands, and the voltage at Bus 58 with and without curtailment.}
\vspace{-3mm}
\label{fig:curtailment}
\end{figure}

The AC OPF formulation was implemented for the distribution network in Orkney, which is constituted of 43 33 kV buses and 30 11 kV buses.
~Fig. \ref{fig:Orkney} presents a simplified version of the assumed network. 
We first run the model for a typical week in June—a period characterised by low demand and, consequently, frequent curtailment events, without incorporating any flexibility from EVs, while maintaining the possibility to curtail wind turbines. This provides a baseline for the amount of curtailment during the period under study. The curtailment events arise when production is considerably above the local demand, although it also depends on the location of the production. 
For the considered week, the overall curtailment reaches 130 MWh, and leads to a regulation of the voltage throughout the network to 1.1 p.u., as shown in Fig. \ref{fig:curtailment} for Bus 58, which is a bus frequently exposed to over-voltages due to its proximity to wind generators.

\begin{figure}[tbp]
\centering
\vspace{-3mm}
\includegraphics[width=0.97\columnwidth]{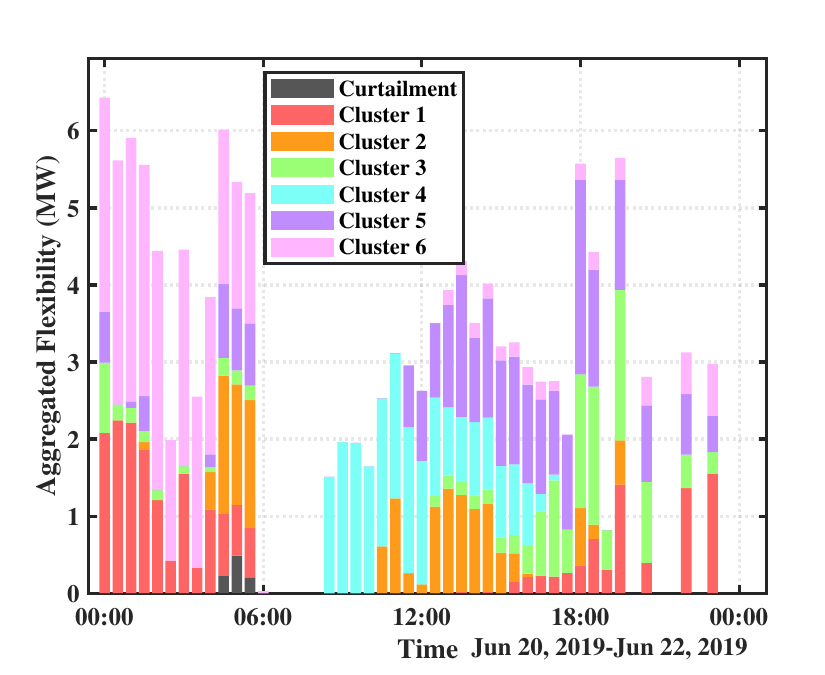}
\vspace{-2mm}
\caption{Flexibility distribution among clusters for the first day of the period considered (for visibility).}
\vspace{-2mm}
\label{fig:results_optim}
\end{figure}

Leveraging the clustering approach defined in Section~\ref{sec:clustering}, next we run the optimisation with EV users' flexibility to reduce curtailment. 
The objective of the optimisation is to replace wind curtailment with flexibility from EV charging. Assuming an EV adoption rate of 80\% (a scenario approaching full electrification of transport in Orkney), i.e. over 13,000 EVs, categorized following the clusters from Fig. \ref{fig:KmeanNEW}, and distributed across the network following the current electric demand distribution, the results are shown in Fig.~\ref{fig:results_optim}.

The results demonstrate that curtailment was reduced by 99.5\% thanks to demand turn-up flexibility provided by EVs that were not already charging, but were assumed to be available for flexibility due to the lower effort required. Importantly, the charging preferences of EV users were respected throughout the process. For instance, Cluster 1, i.e. users who primarily charge their EVs at night, participate predominantly during nighttime curtailment events. In contrast, Cluster 4, which includes users who typically charge during the day, was only active during daytime curtailment events.
This targeted approach enables a more seamless and user-friendly deployment of flexibility, as the algorithm prioritises EV users whose charging behaviour temporally aligns with required flexibility periods. However, in cases of larger wind curtailment events, which require extensive flexibility, all clusters participate in flexibility provision, as shown in Fig. \ref{fig:results_optim} for a major event occurring around 7 p.m., where all clusters except Cluster 4 (whose users never charge in the evening) were engaged. Another important observation is that the required flexible demand power from EVs is usually higher than the optimal curtailment need displayed in Fig. \ref{fig:curtailment}. This is due to the fact that our approach is comprehensive and integrates the grid in the optimisation, leading to larger need of flexible power to overcome network losses and voltage drops, as demand is not always located close to the generation.


\section{Conclusion \& Future Work}
\label{sec:conclusion}

This paper presented a novel AC Optimal Power Flow-based optimisation framework designed to identify optimal flexibility provision from EV owners that are better aligned with flexibility needs, minimising end user disturbance and therefore promoting user acceptance. The proposed formulation ensures that flexibility is only requested from EVs located in grid areas where their contribution is technically effective. Moreover, by incorporating information on EV users’ historical charging habits into the optimisation objective, the approach prioritises users who are more likely to have their EVs connected when support is needed. 
To achieve this, EV users were grouped into clusters with similar charging behaviours using k-means clustering. A novel and improved dimensionality reduction technique based on polar coordinates was developed allowing the method to scale effectively for large datasets containing thousands of EVs.
The method was applied to a real 73-bus distribution network in Orkney, using multi-year charging data from 106 EV users. During a typical summer week, when curtailment is usually highest, the proposed approach with 80\% electrification of transport demonstrated a curtailment reduction of 99.5\%, while only requesting flexibility from EVs within behaviourally optimally aligned clusters.
Future work will incorporate market mechanisms to ensure that EV users are financially compensated for their flexibility services. Additionally, the growing availability of Vehicle-to-Grid (V2G) technologies opens the possibility of extending this framework to manage a broader range of grid constraints, by leveraging connection time information beyond active charging periods.

\section*{Acknowledgment}
This work was supported by EPSRC and UK Department for Transport for Decarbonised Adaptable and Resilient Transport Infrastructures (DARe) project (EP/Y024257/1), the EPSRC “Hydrogen Integration for Accelerated Energy Transitions Hub (HI-ACT)” project (EP/X038823/2), and  DecarbonISation PAThways for Cooling and Heating (DISPATCH) project (EP/V042955/1).

\bibliographystyle{ieeetr}
\bibliography{main_v3}

\end{document}